# Architecting the Future: A Model for Enterprise Integration in the Metaverse

Amirmohammad NATEGHI
*Faculty of Computer Science and engineering*
*Shahid Beheshti University*
*Tehran, Iran*
*am.nateghi@mail.sbu.ac.ir*
*0009-0004-9628-4446*

Maedeh MOSHARRAF
*Faculty of Computer Science and engineering*
*Shahid Beheshti University*
*Tehran, Iran*
*m_mosharraf@sbu.ac.ir*
*0000-0002-6837-7512*
*(Corresponding Author)*

**Abstract— Although it has a history that goes back about three decades, Metaverse has grown to be one of the most talked-about subjects today. Metaverse gradually increased its influence in the realm of business discourse after initially being restricted to discussions about entertainment. Before getting deep into the Metaverse, it should be noted that failure and deviating from the business path are highly likely for an enterprise that relies heavily on information technology (IT) because of improper use and thinking about IT. The idea of enterprise architecture (EA) emerged as a management strategy to address this issue. As the first school of thought of EA, it sought to transform IT from an unnecessary burden in an enterprise to a guiding and supporting force. Then an extended EA model is suggested as a result of the attempt made in this paper to use the idea of EA to steer virtual enterprises on Metaverse-based platforms. Finally, to evaluate the conceptual model and demonstrate that the Metaverse can support businesses, three case studies—Decentraland, Battle Infinity, and Rooom—were utilized.**

*Keywords: Metaverse, Industrial Metaverse, Enterprise Architecture, Virtual Enterprise, Virtual World*

## I. INTRODUCTION

The major definition of an enterprise is to provide services and values for stakeholders, such as customers. Since the expansion of data and information in the enterprise caused complexities and problems in the 1960s, computer systems became a hero to help handle processes and tasks. One particular challenge presented itself in the 1990s: siloed IT systems meant that integration and interoperability appeared as major issues [1]. Once the Clinger-Cohen Law was approved in 1996, enterprises began considering Enterprise Architecture (EA) as a widely recognized approach to seriously aiding the alignment of IT and business objectives [2]. By joining information systems and improving coordination between them, EA helps businesses become less complicated. EA enables an enterprise to experience a variety of advantages, including the foresight to recognize opportunities, the agility to respond to change, and a roadmap to ease the transition to a desired future step [1].

In the contemporary landscape, cutting-edge technologies like cloud computing, big data analytics, Artificial Intelligence (AI), and the Internet of Things (IoT) are fundamentally altering the traditional operational procedures within enterprises and emphasizing the imperative of implementing EA. In essence, these trends serve as catalysts for digital transformation within enterprises. They not only enhance fundamental value and capabilities but also open up a realm of exponentially growing prospects for both services and products [3]. Digitization, as a vital part of EA, enables human beings and autonomous objects to collaborate beyond their working space using digital technologies. The revolution of Industry 4.0 delivered these new technologies, and the collaboration of humans and systems is explained in Industry 5.0, which is a concept of the future of enterprises and industries around a human-centric, sustainable, and resilient system [4, 5].

Metaverse, an innovative technological trend that combines elements from both Industry 4.0 and 5.0, can serve as the cornerstone for a comprehensive digitally-driven business concept. Following Facebook's rebranding to Meta, the Metaverse has become one of the most highly sought-after areas for research and investment[5]. Metaverse is a concept for which there is no common definition. It is a garden in which 100 flowers are blooming and 100 schools of thought contend [6]. However, one of the expressions that introduces the features of the Metaverse well defines it as follows: *"The Metaverse is a massively scaled and interoperable network of real-time rendered 3D Virtual Worlds (VWs) and environments that can be experienced synchronously and persistently by an effectively unlimited number of users with an individual sense of presence and with continuity of data, such as identity, history, entitlements, objects, communications, and payments"* [7]. All the things we have in the actual world could be presented in this VW but in an unreal setting. Users, also known as humans in the Metaverse, can create their arbitrary avatars and carry out regular activities in any Metaverse just as they would in the real world. Fig. 1 depicts the rise of Metaverse acceptance in academic fields in recent years, according to the Google Scholar and DBLP repositories.

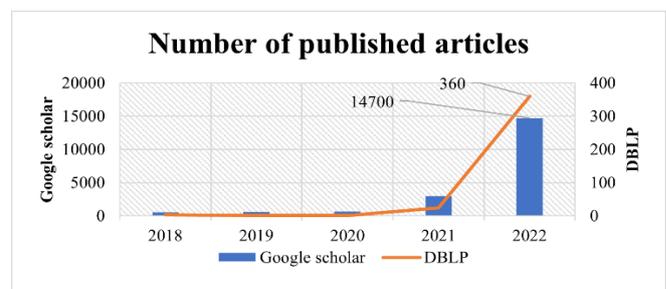

Fig. 1. The number of published articles with the keyword "Metaverse" based on Google Scholar and DBLP (2018-2022)

Changing the needs of enterprises due to the customer's dynamic requests is an inevitable matter that affects the agility and complexity of the enterprises simultaneously. The hardship of overseeing operations, the cost of maintenance, and the difficulties of hiring internationally to be responsive to those demands are instances of influencing factors that encourage the concept of "Virtual Enterprises" (VEs). VE refers to an organization or enterprise that is completely executed based on computer systems. Metaverse can be introduced as a platform for a huge number of VEs with a wide range of goals and purposes thanks to its immersive and three-dimensional capabilities. The translation of this concept into practical applications has only taken place recently; however, the realization of intended enterprises has not been universally successful. This can be attributed to the misalignment between their strategic objectives and the core principles of virtualization. It's worth noting that the establishment or migration of enterprises into the VW, especially concerning staff-related functions, has become a distinct consideration. It appears that the digital transformation journey and the evolution from the fourth to the fifth industrial revolution will make the convergence toward the Metaverse an essential pursuit for all enterprises. This necessity requires a comprehensive reengineering of enterprises and all their objectives. Table. 2 on the next page, which is the result of a study conducted on 39 implemented Metaverse instances, indicates the extent of the enterprise's inclination toward it. As evident from the results of this table, more than half of the Metaverse-implemented platforms encompass the context of business or a combination of entertainment and business. Regardless of platforms, with inaccessible user information, Table 1 illustrates the total number of monthly active users (per 1000) across these platforms categorized by their respective contexts. While the number of registered users in certain VWs in business and multiple contexts is high, it is just one-twentieth of those in the entertainment category. What fuels this pronounced contrast between these segments?

Table 1. Distribution of Metaverse platform users by their contexts

| Context | Total Approximate users (per 1000) |
| --- | --- |
| Business | 6,000 |
| Entertainment | 2,000,000 |
| Multiple | 4,500 |

Given that enterprises weren't initially conceptualized as virtual entities, the evident explanation for this discrepancy is the relative immaturity of enterprises within VWs, despite the long-standing virtual nature of games dating back to the advent of computers. Another rationale for this variation could be attributed to the tendency of enterprises to involve their stakeholders within these virtual environments. This approach diverges from the conventional strategy employed in games, which focuses on maximizing user engagement and encouraging their active participation. EA has emerged as a means to bridge the gap between IT and enterprise operations in our tangible world, akin to the first school of thought [2], [8]. Such an approach seems to be needed in Metaverse enterprises because there are businesses operated over a Metaverse platform, which indicates the remarkable role of IT. There is a significant difference between enterprises in the Metaverse and the physical world. Enterprises in physical mode need EA to improve and stabilize their governance by aligning IT and business concepts, which IT governance will appear as an artifact, but in the Metaverse mode, enterprises try to enhance and stabilize their governance by benefiting from the available IT governance. The intended approach takes the form of a specifically designed EA tailored for Metaverse. This architecture would seamlessly manage the enterprise's business operations within the Metaverse, effectively positioning IT as the foundational pillar of the enterprise's existence. The remaining sections of the paper review the related literature after outlining the background theory of the models used. Then an EA model for the Metaverse enterprises is provided in the following, and finally, it will be reviewed to ensure that it is accurate and in accordance with the relevant enterprises.

## II. THEORETICAL BACKGROUND

Converging EA and Metaverse to craft architectural solutions for Metaverse enterprises demands a comprehensive understanding of both the Metaverse itself, including its architectural structure and established EA models. This section provides an overview of Metaverse architecture and well-regarded EA models.

### A. NIST Enterprise Architecture Model

The following five layers make up the model that the National Institute of Standards and Technology (NIST) has developed to describe the EA idea in Fig. 2:

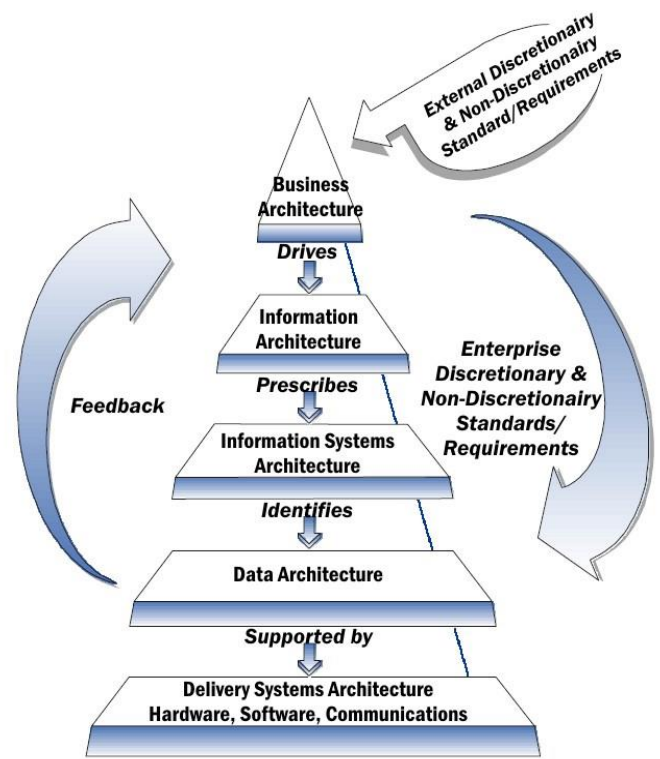

Fig. 2. NIST EA model[9]

- **Business Architecture** refers to managing, navigating, and controlling the design, development, and implementation of the business objectives that are valuable for the company and its stakeholders. This layer demands a high level of analysis and decision-making in the workflow to support and respond to the business mission and vision in addition to objectives [1], [10].

- **Information Architecture** explains the company from the perspective of information and knowledge, which is directly effective on the business layer and its decision-making processes. Information architecture includes identifying and analyzing the information components used by processes within the enterprise. The correlations between various flows of information are also clarified in this architecture, which indicates where, when, and how information is needed and shared to support the responsibility of the functions [10].

- **Information Systems Architecture (also known as Application Architecture)** represents the applications and their interactions in the enterprise. The applications, whether unique to the enterprise or not, are classified based on their designated functions, such as data capture, transformation, management, and storage. However, it's essential to note that this constitutes a broad overview of the applications and their interconnectedness within the enterprise context [1], [10].

- **Data Architecture** illustrates the methods of data maintenance, utilization, and accessibility. Taking a high-level perspective, it establishes and portrays the connections among data components within information systems using three distinct models: conceptual, logical, and physical [1], [10]. Usually, a conceptual model showcases various data entities and their interconnectedness, revealing how they mutually influence one another. This model can be designed either at the level of individual business functions or across entire enterprises, especially when there is a need to represent the information flow in complex supply chains. Generating the conceptual model, a logical data model provides further insights into the data's utilization and comprises three primary elements, including entities, their attributes, and relations. While not mandatory, physical data models can prove to be invaluable initial references for data architects. Leveraging these models, data architects can deduce conceptual data models and, subsequently, logical data models, which form the basis for baseline architectures.

- **Delivery Systems Architecture (also known as Infrastructure Architecture)** is the basis of the EA model, which is known as delivery systems, infrastructure, or technology architecture that supports hardware, software, IT services, and platforms, including the relationships between them [1], [10].

### B. FEAF Framework

Another description of EA is the Consolidated Reference Model (CRM) of the Federal Enterprise Architecture Framework (FEAF), which is an evolved version of the NIST model. This model is specifically tailored for the Office of Management and Budget (OMB) and federal agencies to have a unified language and framework for describing and analyzing investments. In this model, two additional layers named PRM (Performance Reference Model) and SRM (Security Reference Model) have been added to the core five-layer of the NIST model. It's worth noting that the data and information layers in this model are addressed as the DRM (Data Reference Model) layer. The intended model is depicted in Fig. 3.

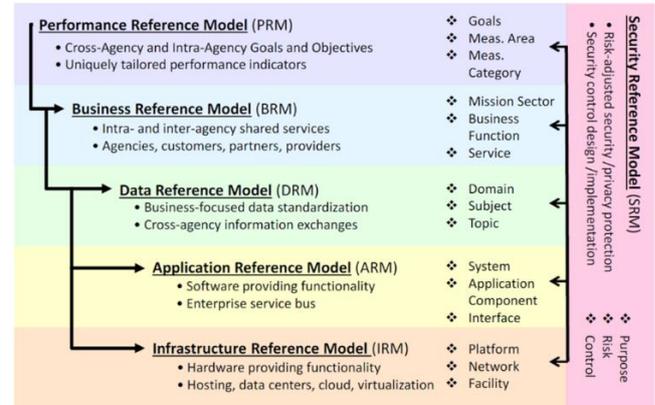

Fig. 3. *CRM from FEAF framework*[11]

- **Performance Reference Model** (PRM) serves as a linkage between enterprise strategy, internal business components, and investments. Its purpose is to provide a measurable way to assess the impact of these investments on strategic outcomes [11].

- **Security Reference Model** (SRM) provides a standardized language and methodology for discussing security and privacy in the context of enterprises' business and performance objectives [11].

In this model, PRM serves as the starting point, connecting the agency's strategic plan to BRM and subsequently to the broader EA. SRM is pervasive, influencing decisions across various sub-architectures to ensure the seamless integration of security into IT systems right from their inception.

### C. Metaverse Architecture

Metaverse is a three-dimensional VW with social interactions, designed to offer a user-friendly interface for vast human interaction and create a comprehensive experience. To integrate the virtual and real worlds in the Metaverse, a combination of various cutting-edge technologies from Industry 4.0, such as networking, AI, IoT, digital twins, blockchain, and XR, is essential. Jon Radoff, the CEO of the Beamable platform, developed a seven-layer model [12] in 2021 to present the architecture of Metaverse, which is shown in Fig. 4.

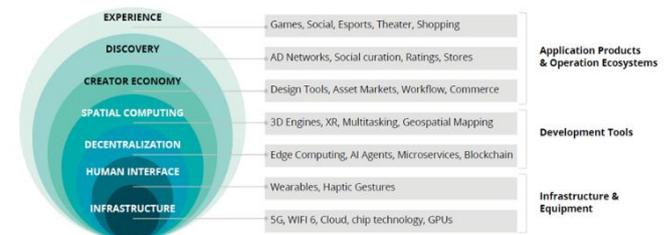

Fig. 4. Architecture of Metaverse[12]

- **Experience:** This layer focuses on providing users with immersive and interactive experiences, enabling them

to engage with VW through various sensory stimuli and interactions. The Metaverse, by removing the constraints of the real world, gives rise to a multitude of personalized services that can offer unique experiences for users. Virtual concerts, where each individual is provided with the best seat, serve as an example of this phenomenon.

- **Discovery:** The discovery layer involves the use of AI and machine learning to assist users in finding relevant content, communities, and activities within the Metaverse.

- **Creator Economy:** This layer empowers users to generate and contribute their content, fostering a creative ecosystem where individuals can share and monetize their creations.

- **Decentralization:** The layer advocates for a decentralized and interoperable structure for the Metaverse, emphasizing experimentation, growth, and sovereignty for creators over their data and creations. Decentralized technologies, such as blockchain, play a crucial role in ensuring transparency, security, and trust within the Metaverse, enabling user ownership and control of their assets.

- **Human-Computer Interaction:** This layer focuses on enhancing the user interface and experience, enabling intuitive and natural interactions between users and the Metaverse. Advancements in technology are evidenced by interfaces ranging from laptops, smartphones, and wireless VR devices to even biological sensors and brain-computer interfaces, underscoring the evolution in this domain.

- **Infrastructure:** The infrastructure layer encompasses the foundational elements that support the entire Metaverse, including networking, data storage, computing resources, and other essential components.

Table 2. The comparison matrix of 39 famous global Metaverses

| Platform | Establishment | Blockchain | Currency | AVG # Monthly Active Users (per 1000) | Context | Country |
|---|---|---|---|---|---|---|
| Second Life | 2003 | - | Linden dollar | 5,000 | E | USA |
| IMVU | 2004 | - | VCOIN | 5,000 | B | USA |
| Roblox | 2006 | - | ROBUX | 1,785,000 | E | USA |
| Minecraft | 2011 | - | Synex coin | 181,227 | E | Sweden |
| Sandbox | 2012 | Ethereum | SAND | 300 | E | Hong Kong, Taiwan and mainland China |
| GTA Online | 2013 | - | In-game Dollar | 15,024 | E | USA |
| Altspace VR | 2013 | - | - | - | - | SHUT DOWN by March 2023 |
| Stageverse | 2017 | Ethereum | ERC721 | - | E | USA |
| Sansar | 2017 | - | Sansar dollar | 6 | M | USA |
| Cryptovoxels | 2018 | Ethereum | - | 3 | B | New Zealand |
| Axie Infinity | 2018 | Ethereum | AXS | 7,782 | E | Vietnam, India |
| Sensorium Galaxy | 2018 | Wakatta | SENSO | - | E | Multi Nation, mainly in the USA |
| WEMIX | 2018 | WEMIX 3.0 | WEMIX | 500 | E | UAE |
| Sorare | 2019 | Ethereum | Own NFT | 650 | E | France |
| GALA | 2019 | Ethereum | GALA | 13,000 | E | Spain |
| Rooom | 2019 | - | - | 310 | B | Germany |
| uHive | 2019 | Ethereum | HVE2 | 400 | B | - |
| Somnium Space | 2020 | Ethereum | CUBE | 21 | E | UK, Canada and Mainly in India |
| Illuvium | 2020 | Ethereum | ILV | 7 | E | Australia |
| Upland | 2020 | EOS | UPX | 150 | E | Ukraine, USA |
| Gather | 2020 | - | GTH | 3,000 | M | USA |
| Decentraland | 2020 | Ethereum | MANA | 240 | B | Argentina |
| Spatial | 2020 | Polygon | GSTA | - | M | USA |
| Microsoft Mesh | 2021 | - | - | - | B | USA |
| Horizon Workrooms | 2021 | - | - | 20 | B | USA |
| Metahero | 2021 | BSC | HERO | - | B | UAE |
| NFT worlds | 2021 | Ethereum | WRLD | 50 | E | - |
| Horizon Worlds | 2021 | - | - | 300 | E | USA |
| Star Atlas | 2021 | Solana | ATLAS | 18 | E | Canada |
| NAKAverse | 2021 | Polygon | NAKA | 200 | E | Thailand |
| High Street | 2021 | Ethereum | HIGH | - | E | - |
| Bloktopia | 2021 | Polygon | Bloktopian | 57 | M | Isle of Man, Ireland |
| Everdome | 2021 | BNB | DOME | - | M | UAE |
| Battle infinity | 2022 | BSC | IBAT | 1500 | M | UK, Canada and Mainly in India |
| Hyper Nation | 2022 | BSC | HNT | - | M | - |
| Viverse | 2022 | Ethereum | - | - | M | Taiwan |
| Efinity | 2022 | Polkadot | EFI | - | M | Singapore |

*E: Entertainment context, B: Business context, M: Multiple contexts*

III.   RELATED WORKS

The background of this research can be examined from both academic and industrial perspectives, which are discussed in separate sub-sections.

*A.  Academic works*

The movement of enterprises toward the Metaverse started with the concept of the industrial Metaverse, which converged with the VE. The industrial Metaverse for smart manufacturing systems is introduced in [13]. This term aims to accelerate various manufacturing processes, such as repairs, starting new manufacturing lines, remote monitoring, and user/manager training through simulation. It utilizes immersive technologies as the configuration layer of cyber-physical systems and acts as a digital twin of the workspace. [14] investigates the expansion of the industrial Metaverse and provides insights into the essential technologies and procedures linked to this groundbreaking development. Furthermore, it explores the ramifications of the industrial Metaverse on various scales, including the firm level, national level, and global level.

[15] has provided an overview of the Metaverse as a potential business platform and its impact on IT industries. It acknowledges that the concept of Metaverse is still unfamiliar to many technical professionals and business managers. The paper proposes a framework for enterprise digitization in the context of the Metaverse. The framework focuses on several critical concepts, such as blockchainization, gamification, tokenization, and virtualization. These concepts are discussed about the 4Ps of the marketing mix: People, Place, Product, and Process.

Several studies within this domain have directed their attention toward the potential impact of the Metaverse on enterprise management processes. [16] proposes a structure for the information space of a VE and develops its core virtual office using augmented reality (AR). It emphasizes the enterprise knowledge base as the main resource for managing the virtual office. Through a simulation experiment on an online store, it was demonstrated that using AR can significantly reduce order processing time, resulting in a threefold increase in efficiency. The authors plan to further develop an integrated virtual environment for virtual enterprises using AR.

The authors in [17] introduce VR-EA+TCK (EA enhanced with Tools, Content, and Knowledge), a solution concept to address the challenges of managing a complex and dynamic IT landscape in EA. Existing digital repositories like Knowledge Management Systems (KMS), Enterprise Content Management Systems (ECMS), and EA Tools (EAT) often remain disconnected, limiting insights and analysis. VR-EA+TCK combines EAT, KMS, and ECMS capabilities in Virtual Reality (VR) to enable stakeholders to visualize and interact with EA diagrams, knowledge chains, and digital entities, promoting grassroots enterprise modeling and collaboration in the Metaverse. The paper presents an implementation and case study to demonstrate the concept's feasibility and potential in various enterprise analysis scenarios. [18] discusses how EA management can support the implementation of VEs for Small and Mid-sized

Enterprises (SMEs). It emphasizes the importance of VEs as a way for SMEs to adapt to new market conditions. The article highlights the significance of managing Information and Communication Technology (ICT) effectively to avoid it becoming a bottleneck in new cooperative efforts. The main focus is on exploring how EA principles and methodologies can enhance the coordination, integration, and alignment of ICT within VEs. The article likely discusses how EA can optimize business processes, information flows, and technology infrastructure to facilitate seamless collaboration and improve overall performance in VEs.

*B.  Industrial works*

The industrial exploration of incorporating Metaverse into enterprise contexts has been dedicated to nurturing an immersive user experience arising from users' active participation within the Metaverse environment—a realm closely mirroring the real world. Industrial enterprises venturing into the use of Metaverse in their operations are pioneers, each with distinct objectives for its utilization. The diverse approaches to employing the Metaverse within the industry can be categorized into the following groups:

- **Production:** The first aspect of production in the Metaverse involves seamless real-virtual coexistence within the research and development phase. Researchers and developers, regardless of their physical locations, can enter the virtual realm together for purposes such as planning, three-dimensional product design, troubleshooting, extensive trial phases, and addressing unstable production processes. The second aspect of real-virtual coexistence in Metaverse production involves managing production through real-time and pervasive simulation of data via IoT and digital twin systems. For instance, companies like Amazon and Pepsi currently utilize similar technologies, such as 5G radio networks and digital twins, to optimize the design of their distribution centers and simulate alternative layout designs [19].

- **Marketing:** While most commercial industries have considered global expansion in their vision, a survey conducted in [20], using data collected from 1015 marketing managers, indicates that even sectors such as energy, resources, industries, life sciences, and healthcare have been inclined toward the Metaverse and are utilizing it for marketing purposes. The companies that use the internet for advertising will also move toward using Metaverse for advertising. For example, while walking in the Metaverse, avatars might encounter advertising billboards. By judiciously integrating Metaverse into the business model, even non-digitally native enterprises can attract younger consumers and update their offerings for the pervasive internet [21].

- **Service Enhancement:** Despite media attention often focusing on revenue potential, some of the most significant applications of the Metaverse within enterprises may lie in achieving equitable access to company processes and developmental opportunities. For instance, various governments like the city of Santa Monica, South Korea, and Saudi Arabia are exploring how the Metaverse can enhance public services [22], [23]. At an enterprise scale, NVIDIA has developed an Omniverse platform where manufacturers like BMW can simulate entire factories. The automaker anticipates about 30% increases in efficiency to optimize floor movements [21].

- **New business models:** Just as some companies in the early 2000s pioneered online-only business models, other companies might also have innovative revenue models built around Metaverse, although they may assume higher risks. Such companies are currently in the process of developing technologies, platforms, products, services, content, and other enabling components of the Metaverse. A notable example is Niantic, the creator of the mobile game Pokémon Go, which provided AR experiences to tens of millions of users, boosting its valuation from $150 million to $9 billion [21]. Snoop Dogg, the renowned rapper, invested $450,000 to acquire a piece of land within the Sandbox Metaverse. He intends to utilize this virtual space for hosting a range of events, including music festivals and concerts, catering to participants seeking immersive experiences in the digital realm [24].

- **Enhanced Workforce Experience:** Several companies are pursuing pervasive technologies like AR and VR to offer personalized experiences for learning and collaboration that are visually engaging, user-friendly, and scalable. These solutions can provide better insights into participation levels, the time apprentices dedicate to lessons, and the stages where they face challenges, thereby enhancing the effectiveness of education. Efforts undertaken by the defense ministry and military forces to create educational environments using digital twins and mixed reality (MR) are important examples of this approach. Soldiers wear glasses that project simulated battle scenarios onto their physical surroundings, preparing them for real engagement situations [25]. The Metaverse also presents a favorable approach for educating high-risk professionals. A notable example is Exelon, the largest electric utility company in the United States, which has gained substantial benefits from incorporating VR training. Given that electrical utility positions can pose dangers to unfamiliar individuals, the virtual environment enables Exelon's employees to engage in learning experiences while wearing protective equipment and solving electrical problems, all without compromising their safety [21].

## IV. INTEGRATING METAVERSE AND EA CONCEPTS

The task of aligning EA concept with Metaverse can be approached from two perspectives, including Metaverse as an enterprise and Metaverse as an infrastructure or virtual platform for implementing an enterprise. Subsequently, we will provide a detailed explanation of each of these proposed approaches.

### A. Metaverse as an Enterprise

Here we have encountered a large-scale enterprise that is still grappling with considerable uncertainties. Aligning EA models with Metaverse, particularly its architecture, presents significant challenges.

- EA is the process through which enterprises align their IT infrastructures with their business objectives. While the business of Metaverse as an enterprise is not yet fully defined, it is still in its infancy. Metaverse, in its role as an enterprise, is far from reaching maturity and currently confronts significant uncertainties.

- The architecture designed for the Metaverse is a combination of various technologies to suit its distinct characteristics. This architecture entails that lower layers are constructed using Metaverse-building technologies. From the standpoint of users, the upper layers of the architecture have been delineated, while considering enterprise dynamics, a synthesis of user, technological, and intermediary layers has been outlined. This is why EA primarily revolves around the enterprise's viewpoint and the integration of IT within its structure. As a result, these two architectures appear to be disparate and not readily comparable.

The CRM model is an enhanced view of the NIST model, which describes an enterprise alignment with the IT concepts in the first school of thought. To more clearly visualize the architecture of the Metaverse-based enterprise, the NIST model was used as the basis and foundation for the concept of EA and CRM as an advanced model of it. As a result, the mapping of various layers of the EA models and the Metaverse architecture model won't be one-to-one. This mapping could take the form of one-to-many, many-to-one, or even occur without a direct counterpart of a layer in the opposing architecture. Fig. 5 illustrates this mapping. As illustrated in this figure, the layer sequence is consistent in both the Metaverse and NIST architectures. The architecture of CRM has been adapted from the NIST model, implying an implicit alignment within it as well. The infrastructure layers refer to all the hardware and network equipment required for the implementation and execution of Metaverse, as well as facilitating user connection, presence, and interaction within it. These devices are equivalent to all the IT infrastructure needed within enterprises. In the Metaverse, the decentralization layer is focused on blockchain infrastructure and maintaining data on a distributed platform. In EA models, this layer can encompass any database schema and data management techniques.

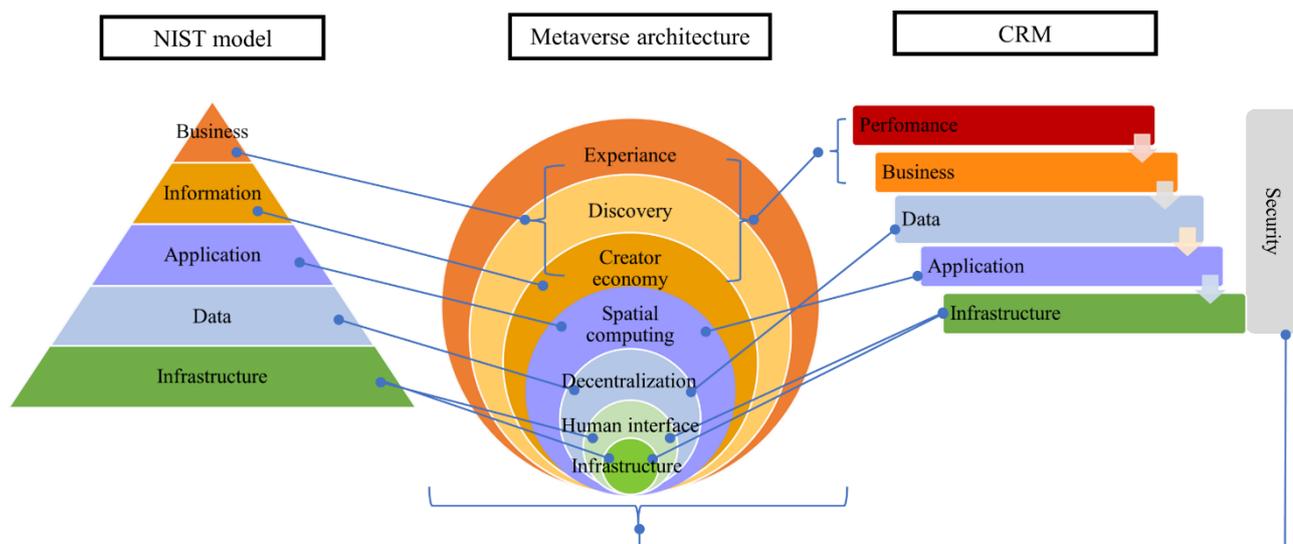

Fig. 5. Incorporating the Metaverse architecture through the EA models

The application layer in the EA models corresponds to the spatial computing layer in the Metaverse model. This alignment is primarily due to one of the focal points of spatial computing, which involves employing digital twin technology for the representation of the real world and mirroring various enterprise use cases within it. However, from another perspective, we can also regard the spatial computing layer as a foundational platform for diverse enterprise applications. In this scenario, platform or user-generated content in Metaverse can shape the desired enterprise applications. Analyzing stored data results in the extraction of information. This process involves processing data stored in the blockchain, thereby establishing the Metaverse economy. The Metaverse economy encompasses the complete financial and digital assets of individuals within this environment.

Integrating the economy, discovery, and experience layers within the Metaverse yields an environment where businesses materialize, processes unfold, and ultimately lead to the acquisition of digital assets or income/tokens. So, the integration of these three layers can seamlessly correspond with the EA business layer. Following this trajectory, the assets that individuals accumulate within the Metaverse, along with their degree of satisfaction within this domain and other measurable factors, can be assessed using the lens of the experience layer. As a result, these three layers seamlessly align with the business and performance layers in the CRM model. Although an enterprise established in the Metaverse might adopt architectures similar to traditional models like NIST or CRM, the complete virtual infrastructure within such entities has necessitated the customization of EA to suit their unique needs. Fig. 6 visually depicts the tailored EA for digitalized or Metaverse-based enterprises.

- **Business Architecture**: As with the origin model, business objectives, processes, and strategy planning are the base concepts of this layer, but in a virtual environment, which seems to be easier, more agile, and less complex. Chief Architecture must mind the demands that are given by the information layer because each decision-making will be effective as soon as possible due to the absence of physical obstacles, which has made it less time-consuming and more flexible. Furthermore, the business services, processes, and tasks within the enterprise are completely transparent and traceable. For instance, in process management, by utilizing the BPMN modeling language [26], there is no longer a need for assembling a human workforce or investing time and expenses into process mining. Incorporating an intelligent system to perform such activities provides process modeling in the shortest time possible and with the highest accuracy.

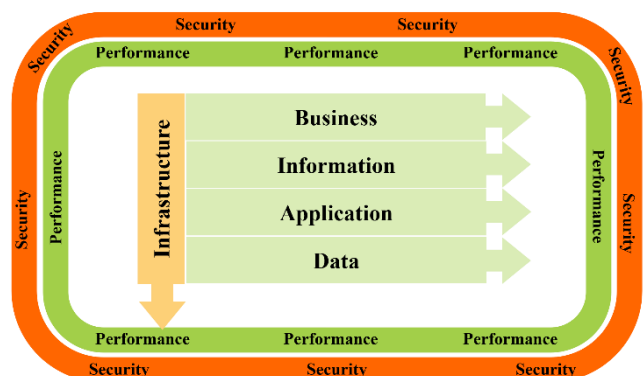

Fig. 6. The proposed model to bring the EA concept through Metaverse-based enterprises

- **Information Architecture**: The essence of the information layer is the same as in the discussed models. Due to full digitalization, a huge mass of information flows can cause the creation of large information silos. Adequate attention is required to sort and classify these information flows to deliver accurate and trusted information to the business layer.

- **Application Architecture**: It is obvious that the application architecture defines and clarifies the relationships between various types of applications. However, a flexible approach to development allows for describing applications as services or microservices within a bank. These can be accessed simply by invoking the designated ones. The important point is that there may be no need to develop and design an application specific to an enterprise.

- **Data Architecture**: In the traditional EA models, data architecture can be described in conceptual, logical, and physical models; these structures are also available here. However, a huge mass of data will be assumed in this layer due to the digitization of all documents and activities, as well as the emergence of a new type of data categorized as personal, behavioral, and communication patterns of enterprise users that must be noted.

- **Infrastructure Architecture**: Infrastructure is now seen as a vertical layer. The reason refers back to the use of technology elements, and the survival of the enterprise depends on the availability of this architecture; without it, the entire enterprise is inaccessible. It's noteworthy that the infrastructure layer still retains its physical nature, as connecting with the Metaverse world might necessitate the use of equipment such as computers, VR headsets, and AR glasses in the most minimal scenario. In physical enterprises, the temporary loss of IT infrastructure results in the cessation of many services and activities (particularly those directly reliant on IT), while some non-IT-dependent activities persist. For example, consider the temporary loss of infrastructure in an in-person educational institution, leading to a slowdown in systemic activities such as grade recording and email communications while teaching and learning continue in a non-systemic manner. However, this scenario doesn't hold for a highly digital enterprise, as the loss of infrastructure in such a case would lead to the enterprise's nonexistence.

- **Performance Architecture**: This layer tries to measure the performance of the inner layers, which is started by determining the Key Performance Indicators (KPIs), measuring the performance, and finding the bottlenecks and Critical Success Factors (CFSs) to handle and design the best architecture development.

- **Security Architecture**: The biggest and most important layer that addresses the security issues in all the main layers; for instance, business security is doing the right processes or infrastructure security as ensuring the CIA triads (Confidentiality, Integrity, and Availability). Another important issue is ensuring data transmission between the real world and the VW is secure, which emphasizes the role of authentication for the users who are attending the specified Metaverse. This is an important gap that was found during the research.

It's important to highlight that the presented model is designed to support Metaverse businesses. By considering Metaverse as an enterprise or a unit of an enterprise, while some businesses can function entirely in a virtual manner, such as educational institutions, others might operate in a semi-virtual mode, like manufacturing companies. So, it doesn't imply that EA should exclusively apply to either fully virtual (EA tailored to VO) or entirely physical enterprises. Taking a broader view, a Metaverse-based company is one place where AI can be widely embraced as an enticing yet somewhat terrifying technology. In every layer of the enterprise architecture, an architect is tasked with designing and structuring the specific layer in alignment with the enterprise's activities and goals. With the full digitization of all data and information, each layer's architecture can be entrusted to one or multiple AI entities known as enterprise architects. Beyond the enterprise architecture, the role of top-level managers can also be entrusted to an AI that plans and defines the enterprise's goals, problem-solving approaches, and pathways to achieve them. If the widespread use of AI occurs significantly, the only human actors within the enterprise might be its customers and owners. This kind of thought expresses an example of a strategy for determining the desired Metaverse's application and future: What is the Metaverse's primary use as a business place, who uses it, and for what purposes?

### B. Metaverse as an Infrastructure

As mentioned in Section 2.1, the infrastructure layer is the foundation of virtual enterprises. Metaverse can be a platform for running a virtual or semi-virtual edition of some enterprises. In this context, IT infrastructures not only align with enterprise strategies but also become essential for the existence of these enterprises. In other words, the Metaverse, as a result of various informational and communicational technologies, is positioned as an infrastructure for all enterprise processes and layers. In this view, Metaverse can be implemented by the enterprise itself or provided as a service by a cloud service provider. Investigating Metaverse as an infrastructure layer in EA models is not included in the scope of this article.

### V. PROOF OF MODEL

As explored in Section 4, if Metaverse is approached as an enterprise, both EA models explored in this paper align well with it. Table 1 presents an examination of around 39 Metaverses, accompanied by their type of data storage. The data underscores a prevalent trend: a significant portion of these Metaverses are structured around distributed databases using blockchain technology. This choice reflects a strategic emphasis on data security within the enterprise layers by harnessing the inherent strengths of blockchain technology. Moreover, capitalizing on the capabilities afforded by this technology, these Metaverses offer a promising vision for both internal and external interactions, enabling users to seamlessly exchange assets across different Metaverse platforms. To assess the applicability of the introduced model to enterprises operating within the Metaverse, an examination of Metaverse-based enterprises becomes imperative. Given the wide array of businesses operating within the Metaverse, this study encompasses three distinct Metaverses representing varying contexts. Table 3 shows the results of applying the Metaverse-based EA model.

- **Decentraland**[1]: Decentraland is a browser-based 3D VW platform that allows users to purchase virtual plots

---

[1] https://decentraland.org/

*Nateghi & Mosharraf*

of land as non-fungible tokens (NFTs) using the cryptocurrency MANA, which is based on the Ethereum blockchain. In this VW, users can interact, explore, and create various experiences on their land. Designers have the opportunity to create and sell virtual clothes and accessories that can be used to customize avatars within the VW. The platform leverages blockchain technology and NFTs to provide a decentralized and unique virtual experience for its users, where ownership of virtual assets is secured and verified through the Ethereum blockchain.

Table 3. Proof of the proposed model with example in the aforementioned Metaverses

| Metaverse \ EA | Decentraland | Battle infinity | Rooom |
|---|---|---|---|
| **Business Architecture** | -Trade anything built by users -Token economy | -New social media -Play P2E -Discover new creations | -Perform workshops, showrooms, etc. shopping in the Metaverse -Monetize virtual or hybrid events -Educations |
| **Information Architecture** | -Monthly active users from specific IP address | -Amount of sold NFTs in a limited timing range | -Amount of held workshops by a specific user |
| **Application** | -Designing tools for personalized assets | -Internal designing tools for NFT designers | -Integrated services to hold conferences |
| **Data Architecture** | -Caught data from the blockchain infrastructure e.g usernames | -Caught data from infrastructure e.g. IP addresses | -Caught data from the blockchain infrastructure e.g. dates |
| **Infrastructure** | -5G networking and its equipment -AR/VR equipment | -5G networking and its equipment -AR/VR equipment | -5G networking and its equipment -AR/VR equipment |
| **Performance** | -Satisfaction of users -Security and privacy | -Customer relationship -Efficiency of infrastructure | -Governance of service -Process effectiveness |
| **Security** | -Availability -Accurate data -Out-of-service applications - Authentication issues | -Amount of attacks -Best routing -Trusted information - Maintaining the financial value of the related token | -Amount of attacks -Defective devices -Out-of-service apps -Wrong business strategies |

- **Battle infinity**[2]: Battle Infinity is a gaming platform that offers a diverse ecosystem of Player-to-Environment (P2E) battle games, all integrated within the VW known as 'The Battle Arena,' which is part of the Metaverse. This platform allows gamers not only to participate in battles and gameplay but also to immerse themselves in the VW of the Metaverse. Within the Battle Arena, users can engage in various activities such as interacting with others, performing actions, watching events, exploring virtual environments, and more. The platform aims to provide a comprehensive and immersive gaming experience within the Metaverse.

- **Rooom**[3]: Rooom is a particular type of enterprise to transform an exhibition, store, or showroom into a virtual space and share with users experiments in digital marketing, sales, education, or events.

Examples of each layer are provided in the above Table to demonstrate that this paradigm is not incompatible with VEs of any type. In contrast to this comparison, EA encompasses a wide variety of documentation, business strategies, artifacts of each layer, EA administration, and others.

## VI. CONCLUSION AND FUTURE WORK

IT is essential to enterprises, perhaps much more so in virtual ones because the entire VE runs on computer systems. Both virtual and physical enterprises can utilize EA to synchronize business management, decision-making, and enterprise vision with IT. This paper broadened the EA idea to handle new issues that virtualization has brought forth because of the endless and pervasive aspects of VWs. The business, information, data, application, and infrastructure layers are the five primary components of this model, and two performance and security layers that surround the main layers complete the entire model, which follows the NIST and CRM models. Although the suggested model's compatibility with the Virtual EA was acknowledged, further evidence is still required to support this claim. As an overview, the first school of thought in EA, which is enterprise IT architecture, has been explained over Metaverse, and the other two thoughts as enterprise integrating and enterprise ecological adaptation, still need more research and analysis [8]. Additional study and analysis are required on documentation, enterprise resource planning, corporate culture, authenticating users as a security feature, and other related topics that can provide useful ideas for future works in the Metaverse context.

### ACKNOWLEDGMENT

None.

### FUNDING

This research received no specific grant from any funding agency in the public, commercial, or not-for-profit sectors.

### AUTHORS` CONTRIBUTIONS

All authors have participated in drafting the manuscript. All authors read and approved the final version of the manuscript. All authors contributed equally to the manuscript and read and approved the final version of the manuscript.

### CONFLICT OF INTEREST

The authors certify that there is no conflict of interest with any financial organization regarding the material discussed in the manuscript.

---

[2] https://battleinfinity.io/

[3] https://www.rooom.com/